# Technical knowledge and soft skills in software startups within the Colombian entrepreneurial ecosystem

Royer David Estrada-Esponda, royer.estrada@correounivalle.edu.co, Universidad del Valle, Colombia
Gerardo Matturro, matturro@fi365.ort.edu.uy, Universidad ORT Uruguay, Uruguay
José Reinaldo Sabogal-Pinilla, jose.sabogal@correounivalle.edu.co, Universidad del Valle, Colombia

**Abstract**: The technical knowledge and soft skills of entrepreneurial team members significantly impact the early stages of software startups. It is widely recognized that the success or failure of a startup is determined by the quality of the individuals who constitute the founding team. This article presents the findings of a study conducted within the Colombian entrepreneurial ecosystem, focusing on which technical knowledge and soft skills are the most valued by founding teams of software startups, and how the needs for knowledge and skills evolve as the startup grows. A survey of software startup representatives revealed that the most valued knowledge includes requirements engineering, software testing, project planning and management, agile methodologies, marketing, business model definition, and budgeting. The most valued soft skills are typically communication, leadership, and teamwork. The outcomes of this work are relevant to software entrepreneurs, incubators, and researchers.

## 1. Introduction

A software startup is a recently created company with little or no operational history, focused on creating and developing an innovative software-intensive product or service as a basis for creating business value (Unterkalmsteiner, 2016), (Giardino, 2014). Among the main challenges of software startups are their scarcity of resources, being highly reactive, being made up of small teams with little experience, relying on a single product, and starting to operate under conditions of uncertainty, rapid evolution, time pressure, and high risks (Giardino, 2014).

These conditions require empirical studies aimed at identifying what software development knowledge and skills are necessary to overcome these challenges, as well as to understand how these knowledge and skills requirements change throughout the evolution of new software companies (Unterkalmsteiner, 2016).

Several studies have shown that the probability of success of a new venture in a dynamic industry increases if the venture includes professionals trained in the various disciplines essential to competing in a complex global economy (Hallam, 2018).

Tanner believes that the success or failure of a business venture is related to the quality of the people who make up the initial team of founders. The new company will flourish or fail depending on how well it recruits, builds, and retains the team (Tanner, 2008), as well as the knowledge, technical abilities, and skills that the members of that team have in relation to the needs and challenges of the venture. A similar opinion is held by Seppänen and colleagues, who believe that it is crucial that a venture obtains the knowledge, abilities, and capabilities necessary to create a product based on innovation (Seppänen, 2017).

This article aims to report on a study focused on identifying the technical knowledge and soft skills most valued in the founding team at the beginning of a software startup, and how these needs evolve as the venture progresses.

The remainder of this paper is organized as follows. Section 2 presents a brief review of the general literature on the concepts of technical knowledge and soft skills and the knowledge and skills considered explicitly for this work. This section also outlines the key characteristics of Colombia's entrepreneurial ecosystem. Section 3 describes this study's methodological design, while Section 4 presents the main findings. Section 5 discusses the results, while Section 6 analyzes the threats to the study's validity. Section 7 outlines some considerations about the relevance of the findings for software entrepreneurs, incubators, and researchers. Finally, Section 8 presents the conclusions.

## 2. Background

This section outlines the key features of Colombia's entrepreneurial ecosystem and provides a brief review of the literature on technical knowledge and soft skills, including the specific types of technical knowledge and soft skills examined in this study.

### 2.1 The Entrepreneurial Ecosystem in Colombia

Colombia's entrepreneurial ecosystem has experienced significant growth in recent years, positioning itself as a vibrant hub for innovation in Latin America. This expansion is fueled by a confluence of several factors, including government initiatives, increasing access to capital, and a burgeoning tech-savvy population.

According to the Colombia Tech Report 2023 (KPMG, 2024), the 1,720 startups across the country included in the report are distributed among 30 sectors; however, 51% of these startups are concentrated in the top six: Fintech (17%), Software as a Service (10%), Business Management (6%), and EdTech, HealthTech, and ProTech, each with 6%, rounding out this top six. Although entrepreneurial activity is widespread across the nation, the primary cities responsible for the most significant volume of startups in Colombia are three: Bogotá (55%), Medellín (25%), and Cali (8%).

Bogotá, the capital, serves as the country's economic and political center, attracting a significant portion of venture capital and fostering a diverse range of startups. Its robust infrastructure, established universities, and concentration of large corporations create a fertile ground for innovation. According to the Global Startup Ecosystem Index 2024 (https://www.startupblink.com/startup-ecosystem/bogota-co), Bogotá boasts the highest-ranked startup ecosystem in Colombia, holds the second position in South America, and is ranked 63rd worldwide. Sectors like fintech, e-commerce, and software development are particularly prominent.

Medellín, once known for its challenges, has undergone a remarkable transformation, emerging as a leading innovation hub. The city's focus on technology and education, epitomized by initiatives like Ruta N (https://rutanmedellin.org/), has significantly contributed to its entrepreneurial resurgence. Medellín's strengths lie in digital media, biotechnology, and advanced manufacturing sectors.

Cali, the third-largest city, is establishing itself as a rising star in Colombia's entrepreneurial landscape. Traditionally known for its agricultural sector, Cali is diversifying its economy, with a growing focus on technology and innovation. Thanks to its reputable universities, the city possesses a strong talent pool, particularly in engineering and computer science.

Colombia's entrepreneurial ecosystem is characterized by a strong sense of community and collaboration. Events and conferences like Open Innovation and Investor Summit Colombia (https://www.oisummit.co/) and the various hackathons held throughout the year provide platforms for entrepreneurs to connect, learn, and pitch their ideas. While each city offers distinct advantages, they collectively contribute to the nation's burgeoning innovation landscape. The government's continued support, coupled with the increasing availability of funding and talent, suggests that Colombia's entrepreneurial ecosystem will continue to flourish in the future.

More details about the Colombian entrepreneurial ecosystem can be found in (Garcia Carvajal, 2022).

**2.2 Technical knowledge and soft skills**

The concept of "technical knowledge" refers to the technical capacity and factual knowledge necessary to do the job and are the technical competencies an individual possesses, acquired through educational learning and its practical application (Bhatnaga, 2012), which are generally associated with the knowledge necessary for the understanding and execution of tasks and processes (Prince, 2013).

On the other hand, the so-called "soft skills" are defined as the combination of skills, attitudes, habits, and personality traits that allow people to perform better in the workplace, complementing the technical knowledge necessary to do their job and influencing the way they behave and interact with others (Matturro, 2019).

**2.3 Knowledge and skills in the founding team**

Entrepreneurs must have "some knowledge" about the industry and markets in which they are involved and about the technology that is relevant to the projected success of entrepreneurial activities. Although entrepreneurs can hire people to fill gaps in their own "skill set", they cannot rely on others for the industry and technology knowledge that is crucial to setting the right course during the entrepreneurial process (Shane, 2003).

When we mention "knowledge in the founding team", we refer to the collective knowledge shared between team members. According to Faulkner (Faulkner, 2022), "collective knowledge" refers to the shared understanding and information that arises from individuals collaborating within a group. It is characterized by the idea that knowledge can be achieved collectively and often depends on contributions from multiple members, as observed in scientific research or team navigation tasks.

**2.4 Technical Knowledge and competencies considered for this study**

Regarding technical knowledge, literature emphasizes that to start and evolve a new venture, the founding team must have specific categories of technical knowledge and competencies, such as management capacity (Riyanti, 2003), (Gianesini, 2018), (Pauceanu, 2026), business competencies (Riyanti, 2003), production capacity (Riyanti, 2003), and financial competencies (Riyanti, 2003), (Pauceanu, 2026), (Gianesini, 2018).

The technical knowledge and competencies categories considered for this study were selected from the referred articles above, and are shown in Table 1.

Regarding "production" knowledge, this refers to the ability to process products, from raw materials to finished products, and includes the technical capabilities required to develop the production process. For software startups, this "production capacity" corresponds to the knowledge necessary for the "creation and development of an innovative software-intensive product or service as a basis for creating business value," according to the definition given at the beginning of this article. With this definition, the technical knowledge in this category was selected from (Cavalcante, 2018), (Klotins, 2019), and (Berg, 2018).

The set of technical knowledge and competencies of the above categories considered for this study is shown in Table 1.

Table 1. Technical knowledge and competencies

| Category | Technical knowledge and competencies |
|---|---|
| Production | Requirements engineering, Prototyping, Systems modeling, User experience design, Architectural design, Coding, Software testing, Wireframing, Software processes, Infrastructure. |
| Management | Agile methodologies, Traditional methodologies, Project planning and management, Quality management, Change management. |
| Business | SWOT analysis, Market analysis, Environmental analysis, Business model definition, Business plan definition, Marketing. |
| Financial | Basic accounting, Budgeting, Cash flows analysis, Investment analysis. |

Tables 2 to 5 describe the meaning of the technical knowledge and competencies elements shown in Table 1.

Table 2. Meaning of Production technical knowledge

| Technical knowledge and competencies Production | Meaning |
|---|---|
| Requirements engineering | The process of defining, documenting, and managing the needs and specifications of a system or product. |
| Prototyping | Creating a simplified early version of software to explore ideas, gather feedback, and refine requirements. |
| Systems modeling | The process of creating abstract representations of a system to understand, design, analyze, or communicate its structure, behavior, and interactions. |
| User experience design | UX design involves creating and optimizing user interactions with a product by understanding needs, behaviors, and preferences. |
| Architectural design | The process of defining a software system's high-level structure involves making key decisions about components, their relationships, and interactions to meet requirements. |
| Coding | Writing instructions in a programming language creates software by translating design specifications and algorithms into code that computers can run, enabling specific tasks. |
| Software testing | The process of evaluating a software application or system to identify defects, verify functionality, and ensure it meets specified requirements. |
| Wireframing | The process of creating a low-fidelity visual representation of a user interface to outline structure, layout, and functionality without focusing on design details. |
| Software processes | Refers to structured activities, methods, and practices used to develop, maintain, and manage software systems throughout their lifecycle, including planning, designing, coding, testing, deployment, and maintenance. |
| Infrastructure | In the context of technology, it refers to the foundational hardware, software, networks, and facilities that support the development, deployment, and operation of software systems and applications. |

Table 3. Meaning of Management technical knowledge

| Technical knowledge and competencies Management | Meaning |
|---|---|
| Agile methodologies | They are iterative, collaborative software development approaches emphasizing flexibility, customer feedback, and continuous improvement. They focus on working software, adaptive planning, and team collaboration over rigid processes, following the Agile Manifesto. |
| Traditional methodologies | Refer to structured approaches like Waterfall and V-model, where phases (planning, design, implementation, testing) are completed sequentially. These methods emphasize documentation, upfront planning, and strict process control. |
| Project planning and management | The process of defining project goals, scope, and deliverables while organizing and coordinating resources, tasks, and timelines to ensure successful software development. |
| Quality management | The process of ensuring products, services, or processes meet standards and customer expectations. It includes quality planning, assurance, control, and continuous improvement to boost efficiency, reduce defects, and maintain consistency. |
| Change management | The process of managing changes to software systems and projects by identifying, documenting, evaluating, and controlling them to minimize disruption, maintain quality, and ensure effective implementation. |

Table 4. Meaning of Business technical knowledge

| Technical knowledge and competencies Business | Meaning |
|---|---|
| SWOT analysis | It is a strategic planning tool used to evaluate the internal and external factors affecting a business or project. It identifies Strengths, Weaknesses, Opportunities, and Threats to provide a clear picture of current conditions and inform decision-making. |
| Market analysis | The process of evaluating a market to understand its size, trends, customer needs, competition, and potential opportunities or risks. |
| Business Environment analysis | The assessment of internal and external factors that impact a business, including economic, political, social, technological, legal, and environmental influences. |
| Business model definition | The process of outlining how a company creates, delivers, and captures value. It describes the key components of a business, including its value proposition, target customers, revenue streams, cost structure, and operational processes. |
| Business plan definition | The process of creating a structured document that outlines a company's goals, strategies, market analysis, financial projections, and operational plan. |
| Marketing | The process of identifying, creating, communicating, and delivering value to meet customer needs and drive business growth. It encompasses market research, branding, product positioning, pricing, promotion, and distribution to attract and retain customers. |

Table 5. Meaning of financial technical knowledge

| Technical knowledge and competencies Financial | Meaning |
|---|---|
| Basic accounting | Refers to fundamental accounting principles and processes used to record and report financial transactions, including tracking income and expenses, managing accounts, preparing financial statements, and ensuring accuracy. |
| Budgeting | Budgeting entails planning and managing financial resources by estimating income and expenses, setting goals, allocating funds, and monitoring spending to control costs, and making informed decisions for short- and long-term objectives. |
| Cash flows analysis | It examines cash movement over a period, tracking inflows from sales, investments, and financing, and outflows for expenses, liabilities, or investments. |
| Investment analysis | It involves assessing financial metrics, market conditions, and economic factors to see if the investment aligns with their financial goals. |

**2.5 Soft skills considered for this study**

Regarding "soft skills", various authors have tried to define and characterize this term. However, the general understanding is that it is a difficult task to accomplish because it is a broad concept, covering many dimensions of personal development and involving a combination of emotional, behavioral, and cognitive components.

Because of this, it is difficult to determine what to include or exclude in its definition. As a way of broadening the base of field study, the soft skills considered for this study were taken from (Komarkova, 2015), (Wrobel, 2018), (Jones, 2021), and (Marr, 2022), and are listed in Table 6.

Table 6. Meaning of soft skills

| Soft skill | Meaning |
|---|---|
| Motivation and perseverance | The ability to stay enthusiastic, determined, and committed to long-term goals despite obstacles. It combines internal motivation with resilience and persistence to keep effort until goals are achieved. |
| Adaptability | The ability to adjust effectively to changing environments, situations, or challenges. It involves being open to new ideas, being flexible in thinking and behavior, and being capable of quickly learning or applying new skills. |
| Learning through experience | The ability to gain knowledge, skills, and insights by reflecting on and analyzing past actions, successes, and mistakes. |
| Leadership | The ability to inspire, guide, and influence others toward shared goals involves effective communication, decision-making, and the capacity to motivate and empower team members. |
| Resilience | The ability to recover quickly from setbacks, adapt to difficult situations, and maintain a positive attitude in the face of adversity. |
| Value creation | The ability to generate ideas, solutions, or innovations that contribute to the success of a project, organization, or customer. |
| Resources utilization | The ability to effectively manage and allocate available resources (such as time, money, materials, or personnel) to achieve goals efficiently. |
| Communication | The ability to clearly and effectively convey ideas, information, and emotions to others. It involves active listening, verbal and nonverbal expressions, and adapting messages to different audiences and contexts. |
| Creative problem solving | The ability to creatively approach challenges by combining analytical skills, adaptability, and innovation to develop solutions. |
| Developing and using contact networks | The ability to build, maintain, and leverage relationships with individuals and organizations to access information, opportunities, and support. |
| Opportunity recognition | The ability to identify and assess potential advantages or gaps can lead to beneficial outcomes. It involves being observant, proactive, and strategic in spotting trends, needs, or problems that can be turned into opportunities for growth, innovation, or improvement. |
| Self-efficacy | Confidence in one's ability to execute tasks, overcome challenges, and achieve goals involves personal competence, resilience, and initiative. |

| Teamwork | The ability to work collaboratively to achieve shared goals by leveraging individual strengths, fostering open communication, and promoting mutual respect. |
|---|---|
| Tenacity | The ability to persist through challenges, setbacks, or obstacles with focus and determination to achieve a goal. It involves resilience, self-motivation, and a strong work ethic, helping individuals overcome difficulties, adapt, and keep striving despite adversity. |
| Convey a compelling vision | The ability to articulate an inspiring long-term goal that motivates and aligns others. It involves persuasive communication to create a shared purpose, helping people see the bigger picture and their role in achieving it. |

## 2.6 Related works

Although the study of general entrepreneurial skills is extensive, very few works were found in the literature referring specifically to the technical knowledge and soft skills most valued for software ventures.

In this area, Santisteban and Mauricio identified, through a systematic review of the literature on success factors in information technology ventures, the following categories of knowledge and skills within the founding team: experience in the industry, academic training, technological and business knowledge, communication and negotiation skills, and leadership (Santisteban, 2017).

Conversely, Seppänen and colleagues focused their study on technical knowledge, highlighting software development and application domain knowledge as the most relevant categories (Seppänen, 2017).

Matturro and colleagues (Matturro, 2020) conducted a study to identify the most valued technical and soft skills within the founding teams of software startups in Uruguay. Through semi-structured interviews with the founding partners from ten software ventures, they discovered that conducting market analysis, defining the business model, and developing the business plan are the most essential technical skills. In contrast, motivation, leadership, and the ability to learn from experience emerged as the most important soft skills. Additionally, as the startups grow, other knowledge and skills, such as financial acumen and securing investments, become increasingly valued, along with soft skills like opportunity recognition and resilience.

## 3 Research Design

The study's methodological design includes formulating research questions, selecting methods and instruments for data collection, and defining data analysis and interpretation procedures.

## 3.1 Research questions

The research questions posed for the study are:

- RQ1: What technical skills are most valued in the founding team of a software startup?
- RQ2: What soft skills are most valued in the founding team of a software startup?
- RQ3: How do the needs for new technical knowledge and soft skills required by the founding team evolve as the venture progresses?

## 3.2 Data collection

Data collection was conducted in two stages, each serving distinct purposes and having unique characteristics, as explained below.

### 3.2.1 Survey

The first stage involved a survey of founding partners or representatives from software startups in various regions of Colombia. This survey aims to gather quantitative data that reflects the diversity of Colombia's entrepreneurial ecosystem in relation to software startups.

For this survey, a questionnaire was developed with a series of closed questions aimed at collecting information relevant to the research questions posed, as well as demographic data from the respondents. The questionnaire was previously validated through pilot interviews with the founding partners of two software ventures.

The survey form was organized into the following sections:

1. **Demographics data**: In this section, respondents are asked whether their startup qualifies as a "software" startup. If so, they are requested to provide additional information, including the startup's name, the year it began operations, the name and email address of the respondent, and their job position within the startup.
2. **Technical knowledge assessment**: This section presents the lists of technical knowledge shown in Tables 2 to 5. Respondents must evaluate each item of knowledge based on a Likert scale with the following options: "Not necessary", "Slightly necessary", "Necessary", "Very necessary", "Essential".
3. **Soft skills assessment**: This section presents the list of soft skills shown in Table 6. Respondents must evaluate each soft skill based on a Likert scale with the following options: "Not necessary", "Slightly necessary", "Necessary", "Very necessary", "Essential".
4. **Evolution of required technical knowledge**: In this section, respondents are asked to differentiate between the technical knowledge they had sufficiently acquired or developed before starting the business and the knowledge they needed to acquire or enhance during its development. "Sufficiently acquired or developed" refers to having a level of knowledge that enables individuals to perform their job adequately and consistently contribute to the company's objectives. The same lists of technical knowledge items shown in Tables 2 to 5 are presented. The answer options were "Before", "During".
5. **Evolution of required soft skills**: In this section, respondents are asked to differentiate between the soft skills they had sufficiently developed at the beginning of the venture and those they needed to acquire or further develop as the startup grows. The same list of soft skills shown in Table 6 is presented, and the options to answer for each one are "Before", "During".
6. **Availability for an interview**: In this final section, we ask the respondent if he/she is available for a follow-up interview on a date to be determined.

To conclude this section, Table 7 illustrates the connection between the research questions posed for this study and the survey sections where the data needed to answer them is gathered.

*Table 7. Relationship between research questions and survey sections*

| Research question | Survey section |
|---|---|
| RQ1 | 2 |
| RQ2 | 3 |
| RQ3 | 4, 5 |

### 3.2.2 Interviews

The second stage of data collection consisted of a series of semi-structured interviews with founding partners or representatives of software startups who had responded to the survey in the previous stage. To achieve this, the last question in the survey inquired whether the respondent was willing to participate in an interview.

The purpose of these interviews was to gather qualitative information about the reasons behind the survey responses, as well as the opinions and perspectives of a small group of interviewees regarding the importance of technical knowledge and soft skills in creating and developing a software startup.

## 4 Results

This section presents the results obtained from analyzing the data collected in the survey. The presentation of these results follows the order of the research questions outlined in section 3.1.

### 4.1 Software startups taking part in the survey

To connect with software startups to participate in the survey, we contacted business incubators and other organizations within Colombia's entrepreneurial ecosystem to ease communication and distribute the online form. This approach allowed us to reach a substantial number of software startups from different regions of the country.

As explained above, to ensure that the responding startups were categorized as "software startups," a discriminatory question (Yes/No) was included at the beginning of the survey. Only if the response was "Yes" would the subsequent survey questions be displayed.

Table 8 shows the geographical distribution of the 74 software startups that ultimately participated in the survey.

*Table 8. Number of participating startups by geographical region*

| Region | Quantity |
|---|---|
| Antioquia | 18 |
| Atlántico | 1 |
| Caldas | 2 |
| Cundinamarca | 22 |
| Quindío | 2 |
| Risaralda | 3 |
| Valle del Cauca | 26 |
| Total | 74 |

Since it is challenging to ascertain the total number of software startups in the country and the number that received the survey form, the survey conducted is considered "exploratory". Conducting an exploratory survey in a scenario where the population size and number of potential respondents are unknown is essential for gaining initial insights about the research topic. This type of survey allows researchers to explore emerging patterns and identify key themes without requiring a predefined sampling frame. Since exploratory surveys do not aim for statistical generalization, they can effectively gather valuable data from available participants, even when the total population is uncertain.

For those who answered the survey on behalf of their startups, we requested them to serve as a founding partner or a representative with a senior management position. This requirement is essential for obtaining responses from individuals with a global perspective on their startups and with "first-hand" knowledge and experience regarding the aspects explored in this study.

Thus, Table 9 shows the distribution of respondents based on their roles in the respective startup.

*Table 9. Distribution of respondents by their roles or positions.*

| Role or position | Quantity |
|---|---|
| Founding partner | 31 |
| CEO | 15 |
| CIO | 11 |
| CTO | 10 |
| Other | 7 |
| Total | 74 |

**4.2 Most valued technical knowledge (RQ1)**

The questions in section 2 of the survey questionnaire enabled us to gather data on the technical knowledge most valued within the entrepreneurial team. To differentiate and rank the relative importance assigned by respondents to each element of technical knowledge, the method employed involved assigning 1 point for each "Necessary" response, 2 points for "Very Necessary," and 3 points for each "Essential" response. Conversely, "Not Necessary" responses received a score of -2, while "Slightly Necessary" responses were assigned -1.

For every knowledge element included in the questionnaire, the total of weighted responses is summed up, according to the following formula:

*Weighted responses = 3\*(Number of answers "Essential") + 2\*(Number of answers "Very Necessary) + 1\*(Number of answers "Necessary) – 1(Number of answers "Slightly Necessary") – 2\* (Number of answers "Not Necessary").*

To establish the ranking, tables were organized from highest to lowest based on the weighted response value. Table 10 presents the ranking of most valued knowledge items in the "Production" category.

*Table 10. Weighted responses for Production knowledge.*

| Knowledge item Production | Weighted responses |
|---|---|
| Software testing | 123 |
| Requirements engineering | 121 |
| Prototyping | 119 |
| Coding | 115 |
| Software process | 104 |
| Systems modelling | 91 |
| Infrastructure | 87 |
| User experience design | 77 |
| Wireframing | 64 |
| Architectural design | 61 |

For the knowledge items in the "Management" category, Table 11 illustrates how respondents ranked them.

*Table 11. Weighted responses for Management knowledge.*

| Knowledge item Management | Weighted responses |
|---|---|
| Project planning and management | 158 |
| Change management | 145 |
| Agile methodologies | 144 |
| Quality management | 110 |
| Traditional methodologies | 62 |

Table 12 presents the ranking of most valued knowledge in the "Business" category.

*Table 12. Weighted responses for Business knowledge.*

| Knowledge item Business | Weighted responses |
|---|---|
| Model business definition | 185 |
| Business plan definition | 178 |
| Marketing | 163 |
| Market analysis | 162 |
| Business Environment analysis | 158 |
| SWOT analysis | 125 |

Finally, in Table 13, the ranking of most valued knowledge in the "Financial" category is presented.

*Table 13. Weighted responses for Financial knowledge.*

| Knowledge item Financial | Weighted responses |
|---|---|
| Budgeting | 160 |
| Investment analysis | 152 |
| Cash flows analysis | 139 |
| Basic accounting | 138 |

**4.3 Most valued soft skills (RQ2)**

Section 3 of the survey questionnaire presented questions about the soft skills most valued within the entrepreneurial team. In all instances, the respondents indicated their preferences using the terms "Necessary," "Very necessary," or "Essential"; the respondents never selected the options "Not necessary" or "Slightly necessary" to refer to these skills. Therefore, to differentiate and assess the relative importance attributed to each soft skill, the weighting method involved assigning 1 point to each "Necessary" response, 2 points to "Very necessary," and 3 points to each "Essential" response.

In this case, the formula for calculating the figures in the "Weighted responses" column is:

*Weighted responses = 3\*(Number of answers "Essential") + 2\*(Number of answers "Very Necessary) + 1\*(Number of answers "Necessary").*

Table 14 presents the ranking of the most valued soft skills, arranged according to the "weighted responses" values.

*Table 14. Weighted responses for most valued soft skills.*

| Soft skills | Weighted responses |
|---|---|
| Communication | 194 |
| Leadership | 191 |
| Value creation | 188 |
| Teamwork | 187 |
| Resilience | 185 |
| Adaptability | 181 |
| Motivation and perseverance | 179 |
| Creative problem solving | 179 |
| Convey a compelling vision | 179 |
| Learning through experience | 174 |
| Opportunity recognition | 173 |
| Developing and using contact networks | 170 |
| Resources utilization | 168 |
| Tenacity | 168 |
| Self-efficacy | 154 |

**4.4 Evolution of technical knowledge and soft skills requirements (RQ3)**

The questions in sections 4 and 5 of the survey gathered the necessary data to address the research question concerning the technical knowledge and soft skills that entrepreneurs typically possess before starting their ventures, as well as those they need to acquire or substantially improve during the venture's development.

For the knowledge elements in the Production category, Table 15 shows the number of responses indicating their acquisition or improvement before the venture began and during its development.

Table 15. Moment of acquiring or improving Production knowledge.

| Knowledge item Production | Before | During | Difference |
|---|---|---|---|
| Coding | 65 | 9 | 56 |
| Software process | 51 | 23 | 28 |
| Requirements engineering | 45 | 29 | 16 |
| Software testing | 41 | 33 | 8 |
| Prototyping | 39 | 35 | 4 |
| Wireframing | 32 | 42 | -10 |
| Systems modelling | 30 | 44 | -14 |
| Infrastructure | 27 | 47 | -20 |
| Architectural design | 26 | 48 | -22 |
| User experience design | 25 | 49 | -24 |

The values in the "Difference" column were calculated as: "Difference = Before – During".

These values in Table 15 indicate that software entrepreneurs are generally well-prepared in coding (+56), software processes (+28), and requirements engineering (+16) before launching their ventures. However, they often need to acquire or significantly enhance their knowledge in user experience design (-24), architectural design (-22), and infrastructure (-20).

The respondents' responses concerning the knowledge elements of the Management category are presented in Table 16.

Table 16. Moment of acquiring or improving Management knowledge.

| Knowledge item Management | Before | During | Difference |
|---|---|---|---|
| Traditional methodologies | 51 | 23 | 28 |
| Project planning and management | 43 | 31 | 12 |
| Quality management | 38 | 36 | 2 |
| Change management | 30 | 44 | -14 |
| Agile methodologies | 28 | 46 | -18 |

According to the meaning of the "Difference" column explained above, most entrepreneurs consider themselves well-versed in project planning and management (+12) as well as traditional methodologies (+28). In comparison, they often need to acquire or improve their knowledge in change management (-14) and agile methodologies (-18).

Table 17 presents the responses regarding the knowledge elements of the Business category.

Table 17. Moment of acquiring or improving Business knowledge.

| Knowledge item Business | Before | During | Difference |
|---|---|---|---|
| SWOT analysis | 44 | 30 | 14 |
| Business model definition | 38 | 36 | 2 |
| Business Environment analysis | 37 | 37 | 0 |
| Business plan definition | 36 | 38 | -2 |
| Market analysis | 34 | 40 | -6 |
| Marketing | 33 | 41 | -8 |

The values in the Difference column of Table 17 indicate that, except for the SWOT analysis (+14), most respondents generally need to acquire or enhance their knowledge in almost all other areas.

Finally, the results of the responses for the knowledge elements in the Financial category are shown in Table 18.

Table 18. Moment of acquiring or improving Financial knowledge.

| Knowledge item Financial | Before | During | Difference |
|---|---|---|---|
| Budgeting | 36 | 38 | -2 |
| Cash flows analysis | 33 | 41 | -8 |
| Basic accounting | 32 | 42 | -10 |
| Investment analysis | 27 | 47 | -20 |

In the Financial category, the values in the Difference column of Table 18 suggest that most entrepreneurs typically need to acquire or improve their knowledge across all areas during the development of their ventures.

Table 19 presents the number of responses based on when the surveyed entrepreneurs acquired or developed their soft skills.

*Table 19. Moment of acquiring or improving the most valued soft skills.*

| Soft skills | Before | During | Difference |
|---|---|---|---|
| Motivation and perseverance | 53 | 21 | 32 |
| Teamwork | 51 | 23 | 28 |
| Communication | 47 | 27 | 20 |
| Adaptability | 44 | 30 | 14 |
| Tenacity | 44 | 30 | 14 |
| Resilience | 43 | 31 | 12 |
| Leadership | 41 | 33 | 8 |
| Self-efficacy | 40 | 34 | 6 |
| Creative problem solving | 38 | 36 | 2 |
| Learning through experience | 37 | 37 | 0 |
| Opportunity recognition | 33 | 41 | -8 |
| Value creation | 33 | 41 | -8 |
| Resources utilization | 32 | 42 | -10 |
| Convey a compelling vision | 30 | 44 | -14 |
| Developing and using contact networks | 27 | 47 | -20 |

Again, based on the values in the Difference column, most entrepreneurs consider themselves well-prepared in the skills of Motivation and Perseverance (+32), Teamwork (+28), and Communication (20) before starting their business. On the other hand, they most often need to acquire or develop the skills of Developing and Using Contact Networks (-20), Conveying a Compelling Vision (-14), and Resources utilization (-10) during venture development.

**4.5 The qualitative interviews**

As explained in Section 3, besides the survey as the primary data collection method, interviews were conducted with some respondents to explore the reasons for their responses and understand their perspectives on the study's objective.

Four of the 74 survey respondents agreed to be interviewed; three are founding partners of their respective software startups, and one serves as the CEO at another software venture.

The interviews were conducted online through Google Meet and lasted an average of 50 minutes. Selected excerpts from the interviews are included throughout the Discussion section to complement and illustrate the analysis of the study's findings with the interviewees' opinions and perspectives.

# 5 Discussion

The key results presented in section 4 are discussed and analyzed in this section.

## 5.1 Overview of the participating software ventures and their representatives

Of the 74 software startups surveyed (Table 8), 89% are in the Antioquia, Cundinamarca, and Valle del Cauca regions. These departments include, respectively, Medellín, Bogotá, and Cali, cities highlighted in section 2.1 as the most dynamic in Colombia's entrepreneurial ecosystem. Although representation from all regions has not been achieved, the three mentioned regions host most of the country's startups.

Regarding the representatives of the ventures that responded to the survey, as shown in Table 9, 90% hold executive roles in their respective ventures, including positions such as founding partner, Chief Executive Officer, Chief Information Officer, or Chief Technology Officer. The "Other" option in Table 9 refers to positions like Team leader, Product Manager, or Sales representative.

The executive roles held by most respondents indicate that they possess enough knowledge and authority to respond appropriately to the survey and to trust the accuracy and relevance of their answers.

## 5.2 On the most valued technical knowledge

From Table 10, four software **production** activities directly related to product engineering emerge as the most valued by software entrepreneurs: requirements engineering, prototyping, coding, and software testing.

These activities form the essential backbone of successful software development. In the dynamic and often resource-constrained environment of software startups, their relevance becomes even more pronounced and critical for survival and growth.

Requirements Engineering in a software startup involves identifying the core value and translating it into a Minimum Viable Product (MVP). Melegati and colleagues (Melegati, 2020) highlight that the idea of "product" is key to understanding MVP among entrepreneurs. They believe MVP is the smallest version with minimal features that provide customer value. Startups must quickly validate ideas, and prioritized requirements ensure initial efforts target the most crucial functionalities for a clear user need.

According to interviewee 4, "*Honestly, we see requirements engineering as highly important because, in the early days, you really don't have a lot of room for guesswork. We needed to be sure we were building something people actually want, and that means taking the time to understand the problem, not just jumping into code.*".

Knowledge about prototyping is exceptionally valuable for an entrepreneurial team in the early stages of a software startup because it provides a rapid and cost-effective way to visualize and test their core ideas before committing significant resources to full-scale development. In words of Interviewee 1, "*Prototyping is central to our process because it allows us to validate our assumptions quickly and cost-effectively. A prototype* *lets us put something tangible in front of potential users or stakeholders early on. It sparks real feedback, not just opinions or guesses.*".

Coding in a startup requires technical skill and speed. Building a solid, scalable system is vital, but startups often need to release a functional product quickly to secure funding, attract users, and stay competitive. This calls for pragmatic coding, emphasizing clean, maintainable code for fast iteration and growth. Poor practices early on can cause technical debt, jeopardizing future development and scalability for a growing startup.

Regarding software testing, it is often seen as a luxury in startups with limited resources, but it is essential to prevent costly failures. Releasing buggy software can damage reputation, erode trust, and risk the startup's survival. Even with limited resources, testing helps find and fix critical defects early, from basic unit tests to user acceptance testing with early users.

Regarding knowledge of **management**, a distinct valuation emerges as shown in Table 15. In the early stages, founding teams tend to place a higher value on knowledge related to project planning and management, agile methodologies, and change management, while seemingly valuing quality management and traditional software methodologies to a lesser extent.

Founding teams are usually small, with members handling multiple roles, focusing on core product delivery over specialized roles or detailed documentation. The main goal is often to find product-market fit amidst uncertainty about the target audience and features. Agile methods are favored for their flexibility, allowing rapid adaptation via short sprints and frequent releases. A study by Mkpojiogu (2019) highlights that key motivations for agile include faster delivery, managing changing priorities, and improving predictability. Strong project planning is essential for defining the MVP and guiding development to validate the product's potential.

When comparing agile approaches to traditional ones, Interviewee 3's opinion is revealing: "*We value agile way more than traditional methods because, honestly, in a startup, things change all the time—your users give unexpected feedback, your priorities shift, you discover better ideas mid-way. Traditional methods just do not handle that kind of flexibility well.*".

The significant importance of knowledge in change management is influenced by the startup landscape's dynamic and often unpredictable nature. In this context, change management knowledge and skills are essential for navigating uncertainties, adapting the product roadmap as necessary, and effectively communicating these changes within the small internal team and to the early user base.

Concerning **business knowledge**, the values in Table 16 show a notable similarity in the high ratings given by respondents. Additionally, relatively higher values are noted compared to those for knowledge of production and management. By placing a high value on business knowledge, the founding teams of software startups acknowledge that their technical (production) knowledge, while essential for product development, is only one aspect of a successful startup equation.

Interviewee 3 mentions, "*Yeah, we put a lot of value on those areas* (business plan, business model, marketing) *because building a great product isn't enough if you don't know how to turn it into a real business.*".

A key reason for focusing on business acumen is to bridge the "product-market fit" gap. Startup teams must validate if their software's problem resonates enough with the target audience to justify payment. Business model definition explains how they will capture value, while marketing helps understand target needs and craft effective messages about their software's value. Regarding marketing, interviewee 2 said "*...that's how we get people to even 'know' we exist. It's what connects us to our users and helps us learn what works and what doesn't.*".

A well-defined business model and thorough market research and analysis can effectively prevent the costly mistake of investing significant time and resources in developing a product that lacks market demand or a clear path to monetization.

This emphasis on business knowledge strategically leverages their technical expertise. A clear business plan highlights opportunities for technical solutions that support growth. In this regard, interviewee 1 stated: "*...we see that business knowledge as just as important as the tech side—because one without the other isn't really a startup, it's just a project.*".

The final set of knowledge elements examined in this study relates to **financial knowledge**, with the respondents' valuations shown in Table 17. This table emphasizes the high valuation of knowledge concerning budgeting and investments.

This prioritization arises from the fundamental need to manage scarce resources effectively and make informed decisions about allocating capital. Early-stage startups often rely on personal savings, angel investors, or venture capital to fuel their initial growth. While their technical skills are crucial for product development, the entrepreneurial teams recognize that without sound financial planning, the startup's runway can be severely limited, hindering its ability to reach key milestones and achieve sustainability. In this regard, interviewee 3 said "*Yeah, we definitely value budgeting and investment knowledge highly, and that's because money decisions can make or break a startup. In the early stage, you're constantly juggling limited resources, so knowing how to budget properly helped us stretch every dollar and stay focused on what really makes the difference.*".

A strong grasp of investment analysis helps the founding team evaluate funding options, negotiate terms, and present a compelling financial case. Pattyn (2023) advises that, with investors becoming more discerning and focusing beyond growth metrics, it's essential to consider metrics like cash flow, ROI, and NPV. Regarding investment analysis, interviewee 4 said "*…on the investment side, understanding how funding works—when to raise, how much to raise, what kind of investors to approach—gives us more control over our future.*".

## 5.3 On the most valued soft skills

Among the soft skills assessed by the survey respondents, four are distinguished as the most valued within the entrepreneurial team, as shown in Table 18: interpersonal communication, leadership, value creation, and teamwork.

**Interpersonal communication** becomes relevant in a usually small founding team. Clear and concise communication minimizes misunderstandings, facilitates the rapid exchange of ideas, and ensures everyone is aligned on goals and progress. Effective communication also extends outwards, playing a vital role in early customer interactions, pitching to potential investors, and building crucial relationships with early adopters and partners.

Regarding **leadership**, even in the absence of formal hierarchies, it is crucial from the outset. Each founding team member needs to take ownership, inspire confidence, and guide others, especially as the team begins to grow. Early leadership sets the tone for the company culture, fosters a sense of shared purpose, and empowers individuals to contribute their best.

The emphasis on **value creation** as a soft skill highlights the importance of the team's ability to collectively understand and articulate the unique benefits their software brings to the market. This goes beyond simply building a functional product; it involves deeply understanding customer needs and translating technical capabilities into tangible value for the end-user.

Finally, **teamwork** is essential in the resource-limited and high-pressure environment of an early-stage software startup. The ability to collaborate effectively, share responsibilities, leverage individual strengths, and support one another through setbacks is crucial for productivity and resilience (yet another highly regarded soft skill).

Regarding communication, leadership, and teamwork, interviewee 2 expressed *"Good communication keeps everyone aligned—no mixed signals, no wasted effort. Leadership, even if it's informal, is what keeps people motivated and focused when things get messy, which they always do. And teamwork? No one's working in a silo in a startup. Everyone's wearing multiple hats, so collaborating, giving and taking feedback, and supporting each other is huge."*.

## 5.4 On the evolution of required knowledge and skills

Startup founders generally possess a strong foundation in core **production** (software engineering) areas before launching, with coding having the highest prior knowledge (Table 15). Teams often also have initial expertise in software development lifecycles, including software process, requirements engineering, testing, and prototyping. However, as ventures mature, there's a significant shift in the knowledge needed, moving from core coding to other areas. Entrepreneurial teams often need to acquire extensive new knowledge during their venture's development, such as user experience design, architectural design, infrastructure, systems modeling, and wireframing. Many teams gain significant understanding in these areas on the job, often surpassing those with prior knowledge.

Regarding **management** knowledge (Table 16), founders often bring foundational management principles into the venture, reporting extensive prior knowledge in traditional areas like project planning and management and traditional methodologies. Despite this, the results show a substantial need for knowledge acquisition during the venture's development. A considerable number of teams had to acquire expertise in all listed management knowledge items while building their startups, indicating that the practical realities and specific demands of a new software venture often necessitate learning on the job.

The landscape for **business** knowledge is mixed, showing both pre-existing expertise and a necessity for on-the-job learning (Table 17). A notable portion of teams possessed extensive knowledge of several fundamental business analysis tools and frameworks before launching, suggesting they bring basic business acumen. Concepts like SWOT analysis and business model definition show a slightly stronger base of prior knowledge, likely because they are foundational concepts taught in business education.

Conversely, a significant trend of acquiring business knowledge during the startup's development is evident, with substantial numbers of teams having to learn and develop their understanding while building their companies. This highlights the nuanced nature of applying business principles in a new software venture. The data suggests a shift towards more learning during the venture for areas directly related to market engagement and strategic planning. For instance, market analysis and marketing show more teams acquiring knowledge during the venture than having it beforehand.

For **financial** knowledge, there's a clear trend where more expertise is acquired during the venture's development than is possessed beforehand (Table 18). Across all listed financial knowledge items, the number of teams that had to learn these skills on their startup journey exceeds the number that had extensive prior knowledge. This suggests that practical financial management in a new software business often necessitates significant on-the-job learning. Areas like cash flow and investment analysis reveal a large gap between prior knowledge and what was learned during the venture, reflecting the complexities of managing finances in a growing startup.

Finally, the survey reveals that many crucial **soft skills** for entrepreneurial success were not well-developed in software startup teams prior to launching (Table 19). The data suggests that skills traditionally associated with individual traits or foundational team dynamics, such as motivation and perseverance, teamwork, and interpersonal communications, are often developed before entrepreneurial pursuits. These skills are built early through education, social settings, and early work experiences. However, the intense demands of a startup environment necessitate further development. The data shows that many teams had to actively acquire or significantly improve skills as their startups grew. The higher numbers for development during the venture for skills like opportunity recognition, value creation, and resource utilization indicate that these competencies become critical as the startup navigates the market and strives for sustainability. A notable crossover point occurs where the number of teams developing a skill during the venture surpasses those who felt it was well-developed before. This shift happens for skills including opportunity recognition, value creation, resource utilization, conveying a compelling vision, and developing and using contact networks.

**6 Threats to validity**

When conducting an exploratory survey where the population size and number of potential respondents are unknown, several threats to validity can arise. These threats impact the study's findings' credibility, reliability, and generalizability.

One of the key threats is **selection bias**. Since our exploratory survey relied on non-probability sampling (e.g., purposive or snowball sampling), the respondents may not represent the broader population. This can lead to biased results, as only certain groups may be included, while others remain unrepresented.

For this study, access to potential respondents was obtained through various business incubators and organizations within the Colombian entrepreneurial ecosystem. However, some software startups may not be associated with these incubators or organizations and, as a result, were not included in the distribution of the survey form.

Another threat corresponds to **sampling bias**. Without knowing the total population size, it is not easy to ensure that the sample adequately represents the diversity of the population. The 74 individuals who responded may possess characteristics or experiences that differ systematically from those who did not participate. The sample might be too small or skewed toward startups who are easier to reach, leading to overrepresentation or underrepresentation of particular perspectives.

A third threat to validity we want to comment on regards to **response bias**. Participants may provide socially desirable answers or be influenced by their personal experiences, rather than reflecting broader trends. In the context of our study targeting software startups affiliated with specific business incubators or organizations, response bias can affect the validity of the study's findings. Since respondents are drawn from specific incubators or entrepreneurial organizations, their views may be shaped by the culture, mentorship, and funding opportunities of those institutions. This can lead to biased responses that reflect the incubator's best practices and philosophies, rather than the broader reality of software startups outside these ecosystems. To address this threat, we tried to collect responses from startups located in various geographical regions of Colombia, and affiliated with different incubators, accelerators, or independent startups to reduce institutional bias.

The last threat we want to reflect on is **external validity**. Findings from our exploratory survey cannot be statistically generalized to a larger population. The primary limitation stems from the unknown number of software startups within Colombia and the potential for a non-representative sampling method. While the survey reached software entrepreneurs in various regions of Colombia, the undetermined distribution method makes it difficult to ascertain if the 74 respondents accurately reflect the diverse landscape of Colombian software startups. Furthermore, the exploratory nature of the survey itself can impact external validity. Exploratory studies often aim to identify initial trends and insights rather than establish definitive, generalizable conclusions.

## 7. Implications for practitioners and researchers

This section outlines key considerations about the relevance of the findings for software entrepreneurs, incubators, and researchers in entrepreneurship and software startups.

The survey results guide future software entrepreneurs by highlighting the most critical technical knowledge for early-stage ventures. Aspiring founders can focus on skill development in requirements engineering, testing, prototyping, agile methods, project management, business modeling, marketing, budgeting, and investment analysis. This approach helps them prepare for initial challenges and increase success by addressing key operational, strategic, and financial areas early. Understanding valued knowledge areas also aids in hiring decisions and identifying where external expertise is needed.

For business incubators, these survey findings offer insights into the needs and gaps of early-stage software startup teams. Incubators can then refine their programs with targeted workshops, mentorship, and resources to develop key technical, managerial, business, and financial skills. This enhances support effectiveness, accelerating startup growth and viability. The findings can also guide selection, prioritizing startups with foundational knowledge or commitment to essential skills, leading to better resource use and outcomes.

Researchers in entrepreneurship and software startups can use these survey results as empirical data to explore the link between specific technical knowledge types and early startup success. The categories and key knowledge items can help develop more detailed research questions and hypotheses on how different knowledge domains affect performance, survival, and growth. This data deepens understanding of essential skills for successful software entrepreneurship, guiding better education, support, and models. Longitudinal studies could examine how the importance of these knowledge areas changes as startups grow.

## 8. Conclusions

In the early stages of software startups, specific technical knowledge and soft skills are especially valued by entrepreneurial teams as they enable startups to move and learn quickly while maximizing limited resources in a challenging business environment.

A survey of founding partners and representatives of software startups from various regions of Colombia identified a set of soft skills and technical knowledge in software production, management, business, and finance that are deemed the most valuable in the early stages of these ventures. The most valued technical knowledge encompasses requirements engineering, software testing, prototyping, coding, project planning and management, change management, agile methodologies, marketing, business model definition, investment analysis, and budgeting. The most valued soft skills include teamwork, interpersonal communication, adaptability, and motivation.

As ventures grow, demand for new or enhanced technical knowledge and soft skills rises. Knowledge and skills that were unavailable or underdeveloped prior to the venture's launch must be obtained or improved by entrepreneurial teams during the venture's growth. Online and in-person courses and hiring experts are preferred methods for acquiring or improving technical knowledge, whereas experiential learning, coaching, and mentoring are favored for developing soft skills.

**CRediT authorship contribution statement**

**Royer David Estrada-Esponda**: Conceptualization, Investigation, Data Curation, Formal analysis, Funding acquisition. **Gerardo Matturro**: Conceptualization, Methodology, Investigation, Formal analysis, Writing - Original Draft, Writing - Review & Editing. **Reinaldo Sabogal**: Conceptualization, Writing - Review & Editing, Supervision, Funding acquisition.

**Acknowledgement**

The authors would like to express their sincere gratitude to the Vicerrectoría de Investigación of Universidad del Valle for their financial support of this research project.